\documentclass[aps,pra,twocolumn,amsmath,showpacs,letterpaper,floatfix]{revtex4}

\usepackage{graphicx}           
\usepackage{booktabs}

\newcommand{\ket}[1]{\left| #1\right\rangle}

\newcommand{\intinf}[0]{\int\limits_{-\infty}^{+\infty}}

\begin{document}

\title{Propagation of Squeezed Vacuum under Electromagnetically Induced Transparency}


\author{Eden Figueroa$^1$\footnote{Present address: Max-Planck-Institut f\"{u}r Quantenoptik, Hans-Kopferman-Strasse 1, D-85748 Garching, Germany.},
Mirko Lobino$^1$, Dmitry Korystov$^1$, J\"urgen Appel$^1$\footnote{Present address: Niels Bohr Institute
Blegdamsvej 17 DK - 2100 K{\o}benhavn  Denmark.} and A. I.
Lvovsky$^1$}

\address{$^1$ {Institute for Quantum Information Science, University of
  Calgary, Calgary, Alberta T2N 1N4, Canada}}
\date{\today}

\begin{abstract}
We analyze the transmission of continuous-wave and pulsed squeezed
vacuum through rubidium vapor under the conditions of
electromagnetically induced transparency. Our analysis is based on a
full theoretical treatment for a squeezed state of light propagating
through temporal and spectral filters and detected using time- and
frequency-domain homodyne tomography. A model based on a three-level
atom allows us to evaluate the linear losses and extra noise that
degrade the nonclassical properties of the squeezed vacuum during
the atomic interaction and eventually predict the quantum states of
the transmitted light with a high precision.
\end{abstract}

\maketitle

\section{Introduction} \label{sec:Introduction}
Electromagnetically-induced transparency (EIT) was first introduced
\cite{harris91:_elect} as a technique to create a transparency
window in an otherwise opaque medium. Years later, Fleischhauer and
Lukin \cite{Lukin_2002} suggested EIT as means for implementing
quantum memory for light, i.e. transferring the state of the
electromagnetic field to a collective atomic spin excitation. After
the first proof-of-principle experimental realization of this
protocol with classical light pulses \cite{lightstorage}, several
experimental groups demonstrated that EIT-based memory can preserve
nonclassical properties of light.

Such experiments became possible with the development of quantum
light sources with wavelength and bandwidth characteristics enabling
strong interaction with the atomic media. Two main approaches have
been pursued. First, the Duan-Lukin-Cirac-Zoller procedure
\cite{Duan20015} has been used to generate narrowband photons which
have then been stored and retrieved using the EIT method
\cite{kuzmichnature, lukinnature}. Second, narrowband squeezed
vacuum tuned to atomic transitions was created by sub-threshold
optical parametric amplifiers (OPAs) \cite{Kozuma_OPO,PKLam_OPO,
Quantech_OPO}, and used to prove the preservation of squeezing after
slow propagation
\cite{Kozuma_EIT_2004,Kozuma_slow_squeezing,Kozuma_cw_SL_EIT} and
storage \cite{Kozuma_storage_squeezing, Quantech_storage_squeezing}.
This last method has also been used recently to preserve the
continuous-variable entanglement in light slowed down by an EIT
medium \cite{PKLam_DelEnt}.

Also, various aspects of the EIT interaction for squeezed light have
been analyzed \cite{Dantan}. Special attention has been paid to the
excess quadrature noise generated by atoms illuminated by the EIT
control field \cite{Peng05,Hetet08}, which was also studied
experimentally \cite{PKLam_noise}.

However, none of the previous publications on slowdown and storage
of continuous-variable optical states reported a satisfactory
comparison of theoretical models with experimental data. In this
paper we fill this void by presenting for the first time a full
theoretical and experimental study of the transmission of squeezed
light through an EIT medium both in the continuous-wave and pulsed
regimes. Our theoretical model uses a series of spectral and
time-domain measurements of the transmission of a classical field
under EIT in order to fit the standard expression for the linear
susceptibility of a 3-level atomic system. Subsequently we use the
susceptibility to calculate the effect of propagation through an EIT
medium of the squeezed vacuum generated by our OPA, both in the
continuous-wave(CW) and pulsed cases. Then we perform two
experimental tests: in the continuous regime, we measure the
quadrature noise spectra of the transmitted squeezed vacuum and, in
the pulsed regime, we perform complete quantum-state reconstruction
via time-domain homodyne tomography. The analysis identifies all
main sources of degradation of squeezing, enabling us to predict the
experimental result with an over 99 \% quantum-mechanical fidelity.

\begin{figure}
\begin{center}
 \includegraphics[width=0.6\columnwidth]{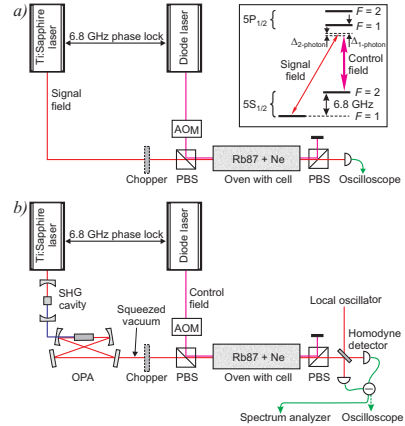}\\
 \caption{Experimental setup for classical (a) and quantum (b) measurements. SHG, second harmonic generation; PBS, polarizing
    beam splitter; AOM, acousto optical modulator. The elements used in experiments with pulsed light are shown with a dashed line.
    The inset displays the atomic energy level configuration.}\label{setup}
\end{center}
\end{figure}

\subsection{Experimental setup}

The EIT medium was an atomic $^{87}$Rb vapor at 65$^{\circ}$C, with
a $\Lambda$ energy level configuration formed by one of the
hyperfine sublevels of the $5P_{1/2}$ state and two hyperfine
sublevels of the $5S_{1/2}$ state (Fig.~1, inset). The classical
signal field, resonant to the
$\ket{5S_{1/2},F=1}$--$\ket{5P_{1/2},F=1}$ transition, was provided
by a Ti:Sapphire laser at a wavelength of 795 nm [Fig.~1(a)]. The
squeezed vacuum signal was obtained from a sub-threshold CW OPA with
a bow-tie cavity, employing a 20-mm-long periodically poled
KTiOPO$_4$ crystal as the nonlinear element \cite{Quantech_OPO}. The
OPA was pumped by the second harmonic of the master Ti:Sapphire
laser and generated a squeezing of approximately 3 dB. The same
laser was used as the local oscillator in homodyne detection
[Fig.~1(b)]. When required, the signal was chopped into 600-ns
(FWHM) pulses using a home-made mechanical chopper
\cite{Quantech_storage_squeezing}. The control field, interacting
with the $\ket{5S_{1/2},F=2}$--$\ket{5P_{1/2},F=1}$ transition, was
obtained from an additional diode laser phase locked at 6.834 GHz to
the master laser to ensure a two-photon resonance with the signal.
Orthogonal, linear polarizations for the control and signal were
utilized. The two fields were focused to a diameter of about 130
$\mu$m and spatially mode matched inside the Rb cell.

\section{Classical measurements}
\begin{figure}
\begin{center}
  \includegraphics[width=0.96\columnwidth]{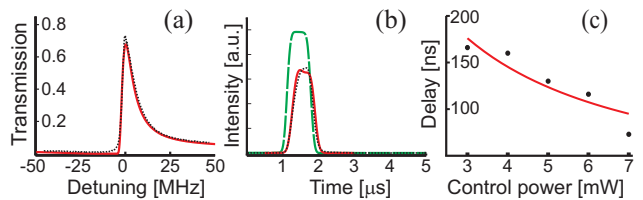}
  \caption {Transmission of classical light through the EIT medium.  a) The EIT
  line transmission spectrum (dotted line) with a theoretical fit (solid line).
  The zero frequency corresponds to the two-photon resonance between the signal
  and control fields. b) Classical input pulse (dashed line), the slowed down
  pulse (dotted line), and a theoretical fit (solid line). c) Time delays for different control
  powers The size of the dot represents the margin of error ($\pm$ 2 ns).
  In (a) and (b), results for 5-mW control field power are displayed.}
  \label{fig:classical_slowdown}
  \end{center}
\end{figure}

The transmission of coherent light through rubidium atoms in the
presence of a control field was measured in order to characterize
the absorption and dispersion of the medium. Assuming a 3-level
system, these properties are governed by the linear susceptibility
\begin{equation}\label{chi}
\chi(\Delta_p,\delta_2)\propto\frac{i\gamma_{bc}+\delta_2}{|\Omega|^2-(i\gamma_{bc}+\delta_2)(\Delta_p+iW)}
\end{equation}
of a Doppler broadened gas under EIT conditions \cite{EIT_decoh}. In
the above equation, $\Omega$ is the control Rabi frequency,
$\Delta_p$ is the one-photon detuning of the signal field,
$\delta_2$ is the two-photon detuning, $\gamma_{bc}$ is the ground
state dephasing rate, and $W$ is the width of the Doppler-broadened
line. The susceptibility determines the transmitted amplitude of the
different frequency components of the input field $E_{in}(\omega)$
through the atomic ensemble by:
\begin{equation}\label{transm_freq}
    E_{out}(\omega)=e^{i\frac{\omega}{2c}\chi L}E_{in}(\omega),
\end{equation}
where $L$ is the medium length, $\omega$ is the angular frequency
and amplitude transmissivity is given by
$T(\omega)=e^{i\frac{\omega}{2c}\chi L}$. The inverse Fourier
transform of Eq. (\ref{transm_freq}) is used in deriving the
temporal profile of the transmitted pulse as shown in Fig
\ref{fig:classical_slowdown}(b) while the time delay was calculated
using the center of mass of the intensity profile.

Equation (\ref{chi}) assumes pure dephasing (decay of the
off-diagonal element of the density matrix) to be the main mechanism
of the ground state decoherence. An alternative decoherence process,
exchange of population between the two ground states, does not
significantly affect the EIT transmission spectrum, as evidenced by
our group's earlier investigations \cite{EIT_decoh} and the quantum
noise measurements described below.

The parameters in Eq. (\ref{chi}) were determined from the following
two measurements, each performed with EIT control powers between 3
and 7 mW. First, the intensity transmission spectrum of the EIT
window was determined by scanning the frequency of the control field
laser while keeping the power and frequency of the signal beam
constant [Fig.~\ref{fig:classical_slowdown}(a)]. Second,
transmission of 600-ns signal pulses was measured in the time
domain, revealing information about the dispersion of the medium
[Fig.~\ref{fig:classical_slowdown}(b)]. As expected from an EIT
system, significant (up to a factor of a thousand) reduction of the
group velocity of light has been observed
[Fig.~\ref{fig:classical_slowdown}(c)].

The results were fitted with Eq. (\ref{chi}). The data for different
control powers were fitted with the same set of parameters, except
the control Rabi frequency, which was proportional to the square
root of the control field intensity. Although treating the rubidium
atom as a three-level system is an oversimplification, it turns out
sufficient to obtain good agreement, as evidenced by
Fig.~\ref{fig:classical_slowdown}.

\section{Measurements on squeezed vacuum}
\subsection{Squeezed vacuum: CW measurements}
\subsubsection{OPA squeezing}
In order to predict the evolution of squeezing under EIT conditions,
we need to revisit the classic theory of squeezing inside an OPA
cavity as developed by Gardiner and Savage \cite{GardinerSavage} and
reviewed, for example, in Ref.~\cite{Scully}.

In a degenerate, continuous-wave OPA pumped below threshold at
frequency $2\Omega$, parametric-down conversion produces pairs of
photons at lower frequency $\Omega\pm\omega$. In the Heisenberg
picture, it can be described as nonlinear mixing of input vacuum
fields $a_{\rm in}(\pm\omega)$ associated with optical frequencies
$\Omega\pm\omega$ (Bogoliubov transformation):
\begin{equation}\label{OPOout}
\hat a(\omega)=C(\omega)\hat a_{\rm in}(\omega)+S(\omega)\hat a_{\rm in}^\dag(-\omega),
\end{equation}
where, under certain approximations,
\begin{eqnarray}\label{CS}
  C(\omega) &=& 1-2\gamma\frac{\gamma - i\omega}{(\gamma -i\omega)^2-\gamma^2 P/P_{th}};\\ \nonumber
  S(\omega) &=& 2\gamma^2\frac{\sqrt{P/P_{th}}}{(\gamma -i\omega)^2-\gamma^2P/P_{th}}.
\end{eqnarray}
In Eq. (\ref{CS}), $P/P_{th}$ is the ratio of the OPA pump and
threshold powers, and $2 \gamma$ is the cavity bandwidth. Using
$[a_{\rm in}(\omega),a^\dag_{\rm
in}(\omega')]=\delta(\omega-\omega')$, we find the correlations of
the OPA output fields at different frequencies:
\begin{eqnarray}\label{acorr}
\nonumber
  \langle\hat{a}^{\dag}(\omega)\hat{a}(\omega')\rangle &=& S^*(\omega)S(\omega')\delta(\omega-\omega'); \\
  \langle\hat{a}(\omega)\hat{a}(\omega')\rangle &=& C(\omega)S(\omega')\delta(\omega+\omega').
\end{eqnarray}

The frequency-domain quadrature observable associated with phase
$\vartheta$ is defined as \cite{footnote0}
\begin{equation}\label{quadrop}
\hat{q}_{\vartheta}(\omega)=\frac 1{\sqrt 2}\left[\hat{a}(\omega)e^{i\vartheta}+\hat{a}^{\dag}(-\omega)e^{-i\vartheta}\right].
\end{equation}
The phase-dependent quadrature noise is then given by
\begin{eqnarray}\label{eq.variance_quadrature_fourier4}
\nonumber\langle\hat{q}_{\vartheta}(\omega)\hat{q}_{\vartheta}(-\omega')\rangle &=&\frac{1}{2}
 \langle
    \hat{a}(\omega)\hat{a}^{\dag}(\omega')+\hat{a}^{\dag}(-\omega)\hat{a}(-\omega') \\ \nonumber
    &&\hspace{-1cm}+\ \hat{a}(\omega)\hat{a}(-\omega')e^{+2i\vartheta}
    +\hat{a}^{\dag}(-\omega)\hat{a}^{\dag}(\omega')e^{-2i\vartheta}
 \rangle\\
 &&\hspace{-1cm}\equiv\ V_\vartheta(\omega)\delta(\omega-\omega'),\end{eqnarray}
where
\begin{eqnarray}\label{eq.variance_quadrature_fourier_CD}
 V_\vartheta(\omega)&=&\frac{1}{2} \large[1+S^*(\omega)S(\omega)+S^*(-\omega)S(-\omega)\\ \nonumber
&+&C(\omega)S(-\omega)e^{+2i\vartheta}+C^*(\omega)S^*(-\omega)e^{+2i\vartheta}\large].\nonumber
\end{eqnarray}
Analyzing Eq. (\ref{CS}), we find that $S^*(\omega)=S(-\omega)$ and
$C(\omega)S(-\omega)$ is a real number for all frequencies, which
allows the above expression to be simplified:
\begin{equation}\label{eq.variance_quadrature_fourier_CD1}
 V_\vartheta(\omega)=
\frac 1 2+|S(\omega)|^2+
C(\omega)S(-\omega)\cos(2\vartheta).
\end{equation}
In order to relate this quantity to the noise spectrum measured
experimentally by means of homodyne detection, we need to account
for imperfect quantum efficiency of the photodiodes and the optical
losses inside and outside the OPA cavity. Equation
\eqref{eq.variance_quadrature_fourier_CD1} then becomes
\begin{equation}\label{eq.variance_quadrature_fourier_CD2}
 V^{\rm exp}_\vartheta(\omega)=
\frac 1 2+\eta[|S(\omega)|^2+
C(\omega)S(-\omega)\cos(2\vartheta)],
\end{equation}
where $\eta$ is the overall quantum efficiency. Substituting Eq.
(\ref{CS}), we obtain the well-known result for the quadrature noise
levels associated with phases $\vartheta=\pi/2$ (antisqueezed) and
$\vartheta=0$ (squeezed):
\begin{equation}\label{OPAout}
V^{\pm}(\omega)=\frac 1 2 \pm\eta\frac{2\sqrt{P/P_{th}}}{(\omega/\gamma)^2+(1\mp\sqrt{P/P_{th}})^2}.
\end{equation}

Experimentally, we observe continuous-wave squeezing in the
frequency domain by feeding the homodyne detector output to a
spectrum analyzer. At each frequency, the local oscillator phase was
varied and the highest and lowest noise levels were recorded. Figure
\ref{fig:CWsqueezing}(a) shows the experimentally measured highest
and lowest quadrature noise levels of the OPA output and a
theoretical fit with Eq. (\ref{OPAout}).

\subsubsection{Propagation through the EIT medium}
\begin{figure}[t]
\begin{center}
  \includegraphics[width=0.6\columnwidth]{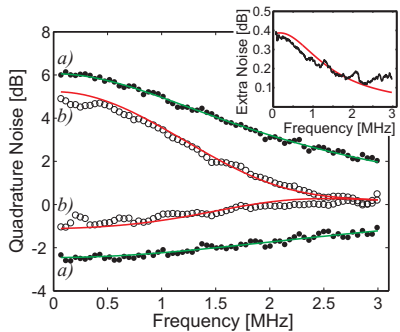}
  \caption{Squeezed and antisqueezed quadrature noise from the
  OPA (a) and after the EIT medium (b), along with the theoretical prediction of Eq.~(\ref{OPAout}) and Eq.~(\ref{eq.CWsqueezing}) plus the extra noise.
  Inset: atomic extra noise measured in the absence of the signal field. The control field
  power is 5 mW. The error bar is $\pm0.05$ dB and is smaller then the size of the dots.}
  \label{fig:CWsqueezing}
 \end{center}
\end{figure}

We have placed the atomic vapor cell between the OPA and the
homodyne detector and performed the frequency-domain measurement of
the quadrature noise of the transmitted squeezed vacuum. We found
that the squeezing is preserved better when the lasers are slightly
detuned from the two-photon resonance. This is because the noise
reduction observed at a specific electronic frequency $\omega$
originates from quadrature entanglement between the spectral modes
associated with the frequencies $\Omega\pm\omega$ of the optical
signal field. Thus both modes need to be transmitted through the EIT
transparency window. Because our EIT lines are substantially
asymmetric [Fig. 2(a)], one of the modes is strongly absorbed under
the conditions of two-photon resonance. Figure
\ref{fig:CWsqueezing}(b) displays CW squeezing that remains after
the interaction with rubidium atoms with a two-photon detuning of
540 kHz.

In our theoretical analysis, we assume two main degradation
mechanisms. First, there is linear loss due to a finite width and
imperfect transmission of the EIT line. Upon propagation through a
medium with amplitude transmissivity $T(\omega)$, the field operator
becomes $\hat a'(\omega)=\sqrt\eta T(\omega)\hat
a(\omega)+\sqrt{1-\eta|T(\omega)|^2}\hat v(\omega)$, where $\hat
v(\omega)$ denotes vacuum. In writing this equation we have assumed
that the losses associated with the quantum efficiency $\eta$ occur
before the rubidium cell; assuming otherwise would not change the
results below since only linear losses are involved. Substituting
the new field operator into Eq.
(\ref{eq.variance_quadrature_fourier4}) and noticing that $\hat
a(\omega)$ and $\hat v(\omega)$ are uncorrelated, we find for the
quadrature noise
\begin{eqnarray}\label{eq.CWsqueezing0}
 V'_\vartheta(\omega)&=&\frac 1 2+\frac{\eta}{2} \large[|S(\omega)|^2[|T(\omega)|^2+|T(-\omega)|^2]\\ \nonumber
&+& C(\omega)S(-\omega)[T(\omega)T(-\omega)e^{2i\vartheta}+T^*(\omega)T^*(-\omega)e^{-2i\vartheta}]\large].
\end{eqnarray}

From Eq. (\ref{eq.variance_quadrature_fourier_CD2}) we obtain
\begin{eqnarray}\label{CSexp}
\eta|S(\omega)|^2&=&\frac{V^+(\omega)+V^-(\omega)-1}2;\\ \nonumber
\eta C(\omega)S(-\omega)&=&\frac{V^+(\omega)-V^-(\omega)}2,
\end{eqnarray}
which allows us to relate the maximum and minimum quadrature noise
levels in the presence and absence of the vapor cell:
\begin{eqnarray}\label{eq.CWsqueezing}
    V'^\pm(\omega) &=& \frac 1 2+ \frac 1 2
    \left[V^+(\omega)+V^-(\omega)-1\right]\\
    &\times&\left[|T(\omega)|^2+|T(-\omega)|^2\right]\nonumber\\
    &\pm&\left[V^+(\omega)-V^-(\omega)\right] |T(\omega)| |T(-\omega)|\nonumber.
\end{eqnarray}
Interestingly, for each detection frequency this noise depends only
on the magnitude of $T(\omega)$, but not on the phase shift
introduced by the rubidium gas. This is because any complex phase in
the quantity $T(\omega)T(-\omega)$ in Eq. (\ref{eq.CWsqueezing0})
that may be present due to asymmetric dispersion of the EIT line can
be compensated by adjusting the local oscillator phase $\vartheta$.

The degradation of squeezing in transmission through EIT occurs not
only due to absorption but also because of extra noise  $V_{\rm
noise}(\omega)$ generated by atoms in the cell \cite{PKLam_noise}.
This contribution is a consequence of the population exchange of the
rubidium ground states and has been investigated theoretically in
Refs.~\cite{Peng05,Hetet08}. We evaluated this noise independently
by performing homodyne detection of the field emanated by the EIT
cell in the absence of the input squeezed vacuum. The result of this
measurement, with the shot noise subtracted, is shown in the inset
of Fig.~\ref{fig:CWsqueezing}, along with a theoretical fit
according to Ref.~\cite{Hetet08}. The only additional fitting
parameter here is the population exchange rate, which determines the
overall magnitude of the extra noise \cite{footnote}. We estimate
this rate to account for less than 10 \% of all decoherence
processes, which justifies our earlier assumption.

We account for the extra noise by adding $V_{\rm noise}(\omega)$ to
the right-hand side of Eq. (\ref{eq.CWsqueezing}). In this way, we
construct a theoretical prediction for the spectrum of the
transmitted squeezed vacuum. No additional fitting parameters are
required. The agreement between theory and experiment in
Fig.~\ref{fig:CWsqueezing}(b) shows that, indeed, absorption and the
extra noise are the main factors responsible for the degradation of
squeezing.

\subsection{Squeezed vacuum: pulsed measurements}
\begin{figure}[b]
\begin{center}
    \includegraphics[width=0.7\columnwidth]{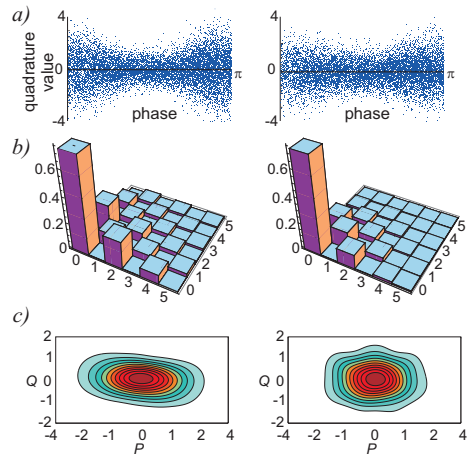}
  \caption{Quantum state of the input (left column) and slowed down (right column) squeezed vacuum with an EIT control power of 5 mW.
  Raw samples of phase-dependent quadrature noise (a), maximum-likelihood reconstruction
  of the density matrices in the Fock basis (absolute values, b) and Wigner functions (c) are shown.}\label{fig:Tomography}
  \end{center}
\end{figure}

\begin{figure}
\begin{center}
 \includegraphics[width=0.6\columnwidth]{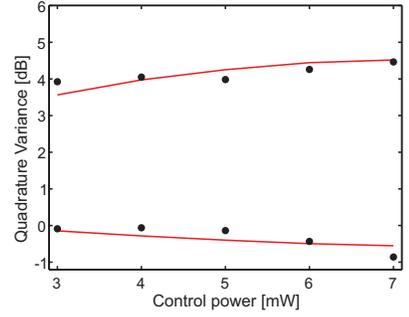}
 \caption{Maximum and minimum quadrature noise of pulsed squeezed vacuum
 transmitted through the EIT cell (theory and experiment). The size
 of the dot represents the margin of error ($\pm$ 0.05 dB).}
 \label{fig:pulsedsqueezing}
 \end{center}
\end{figure}

We studied propagation of pulses of squeezed vacuum through the EIT
medium by means of time-domain homodyne tomography, using a
procedure similar to that of Ref.~\cite{tempfiltering}. The output
of the homodyne detector was multiplied by the temporal mode
function $W(t)$ and integrated over time, producing a single sample
of the field quadrature. The weight function was equal to the square
root of the transmitted classical pulse intensity [dotted line in
Fig.~\ref{fig:classical_slowdown}(b)]. In this manner, 50,000
quadrature samples were obtained. The local oscillator phase values
associated with each quadrature were determined using the method
described in Ref.~\cite{Quantech_storage_squeezing}, by measuring
the degree of squeezing in the transmitted light with a spectrum
analyzer running continuously. These data were used to reconstruct
the density matrix and the Wigner function of the transmitted,
pulsed optical mode by means of the maximum-likelihood
algorithm\cite{Lvovsky_MLR} (Fig.~\ref{fig:Tomography}). The
reconstructed optical state strongly resembles a squeezed thermal
state. The minimum and maximum noise levels observed at different
control field powers are summarized in
Fig.~\ref{fig:pulsedsqueezing}.

The experimental data on classical transmission of the EIT line, the
parameters of the OPA and the atomic extra noise spectra are
sufficient to make theoretical predictions for the transmission of
squeezed vacuum in the pulsed regime without additional fitting
parameters.

We now begin the analysis in the time domain. Suppose the output of
the OPA cavity is expressed by a time-dependent annihilation
operator
\begin{equation}\label{aoft}
 \hat{a}(t)=\frac 1 {\sqrt{2\pi}}\intinf\hat a(\omega)e^{-i\omega t} {\rm d} \omega.
\end{equation}
After the chopper, this operator will change into
$\hat{a}'(t)=\sqrt\eta\tau(t)\hat{a}(t)+{\rm v.c.}$ with $\tau(t)$
representing the time-dependent transmission of the chopper and {\rm
v.c.} denoting the vacuum contribution whose explicit form will be
calculated later. In order to account for the propagation through
the cell, we switch to the frequency domain, replacing
multiplication by convolution:
\begin{equation}\label{aprimfreq}\hat{a}'(\omega)=\sqrt\frac \eta {2\pi}\intinf\tilde{\tau}(\omega-\omega')\hat{a}(\omega'){\rm d}\omega'+{\rm v.c.},
\end{equation} where $\tilde{\tau}(\omega)$ is the Fourier images of the $\tau(t)$. Absorption in the rubidium cell is then described by
\begin{equation}\label{freqfilter}
\hat{a}''(\omega)=T(\omega)\hat{a}'(\omega)+{\rm v.c.}
\end{equation}
After the cell, the quantum state of the temporal mode defined by
the function $W(t)$ is subjected to homodyne tomography. The field
operator associated with this mode is given by
\begin{eqnarray}\label{tempmode}
\hat A&=&\intinf \hat a''(t) W(t){\rm d}t=\intinf \hat a''(\omega) W(-\omega){\rm d}\omega\\ \nonumber
      &=&\intinf F(\omega) \hat a(\omega) {\rm d}\omega + \intinf G(\omega) \hat v(\omega) {\rm d}\omega,
\end{eqnarray}
where
\begin{equation}\label{Comega}
F(\omega)=\sqrt\frac \eta {2\pi}\intinf T(\omega')W(-\omega')\tau(\omega'-\omega){\rm d}\omega',
\end{equation}
$\hat v(\omega)$ is the vacuum field operator and the weight of the
vacuum fraction is given by $G(\omega)=\sqrt{1-|F(\omega)|^2}$. The
measured (time-integrated) quadrature is given by $\hat Q=(\hat
Ae^{i\vartheta}+\hat A^\dag e^{-i\vartheta})/\sqrt 2$ and its
variance, according to Eqs.~(\ref{acorr}) and (\ref{tempmode}) is

\begin{eqnarray}\label{eq.pulse_var_final0}
   V''_{\vartheta}&=&\langle \hat Q^2\rangle=\frac{1}{2}+V_{\rm noise}+\intinf |F(\omega)|^2|S(\omega)|^2\\ \nonumber
   &+&\frac 1 2\left[\intinf C(\omega)S(-\omega)F(\omega)F(-\omega)e^{2i\vartheta}\ +\ {\rm c.c.}\right] ,
\end{eqnarray}
which using Eq. (\ref{CSexp}) simplifies to
\begin{eqnarray}\label{eq.pulse_var_final}
   &V&''_{\vartheta}=\frac 12+V_{\rm noise}\\
   &+&\frac{1}{2}\int_{-\infty}^{+\infty} |F(\omega)|^2\left[V^+(\omega)+V^-(\omega)-1\right]d\omega\nonumber\\
    &+&\frac{1}{2}\int_{-\infty}^{+\infty} |F(\omega)F(-\omega)|\left[V^+(\omega)-V^-(\omega)\right]\cos(2\vartheta+\varphi(\omega))d\omega  \nonumber
\end{eqnarray}
with $\varphi(\omega)={\rm arg}[F(\omega)F(-\omega)]$. The pulsed
atomic extra noise is obtained from the frequency-domain extra noise
by integrating
\begin{equation}\label{Vnoise} V_{\rm noise}=\int_{-\infty}^{+\infty}|W(\omega)|^2 V_{\rm noise}(\omega)d\omega.
\end{equation}

Unlike the CW case, the phase shift imposed by the EIT line does
lead to additional loss of squeezing. Because the EIT line is
asymmetric [i.e. $T(-\omega)\ne T^*(\omega)$] and $\varphi(\omega)$
is not constant, the squeezed and antisqueezed quadratures may mix
with each other in the integral \eqref{eq.pulse_var_final}.

The solid line in Fig.~\ref{fig:pulsedsqueezing} shows the result of
the calculation. For all control powers, over 99\% fidelity (defined
as $F={\rm
Tr}[(\hat\rho^{1/2}_{th}\hat\rho_{meas}\hat\rho^{1/2}_{th})^{1/2}]^2$)
between the theoretically predicted and experimentally observed
states is reached.

\section{Summary}
We have performed a theoretical and experimental investigation of
squeezed light propagation through an EIT medium in both the CW and
pulsed regime. Starting with a theoretical expression for the
susceptibility of the EIT medium, we determined the degradation of
squeezing in each spectral component of the squeezed vacuum. For the
pulsed case, we performed full quantum reconstruction of the optical
state after propagation through the rubidium cell and compared it
with the one predicted theoretically. We identified all main
mechanisms leading to the degradation of squeezing: absorption in
the EIT medium, asymmetry of the EIT line, and the extra noise
induced by the control field. A very simple 3-level model is
sufficient to fully explain our experimental results. This latter
conclusion confirms a recent result of classical experiments on
storage of light \cite{Irina}.

\section{Acknowledgements}
We gratefully acknowledge F. Vewinger, K. P. Marzlin and B. Sanders for fruitful
discussions and C. Kupchak for assistance in the laboratory.
This work was supported by NSERC, CIAR, iCORE, AIF, CFI and Quantum\emph{Works}.

\section*{References}


\begin{thebibliography}{13}
\bibitem{footnote0}
Note that $\hat{q}_{\vartheta}(\omega)$ is, generally speaking, not Hermitian, but $\hat{q}_{\vartheta}(\omega)=\hat{q}_{\vartheta}^\dag(-\omega)$.
\bibitem{footnote}
The theory of Ref.~\cite{Hetet08} was developed for the case of zero one-photon detuning, i.e. symmetric EIT line. We used this approximation when constructing the fit.
\end{thebibliography}
\end{document}